# Lattice-polarization effects on electron-gas charge densities in ionic superlattices

D. R. Hamann,[1,2,3] D. A. Muller,[4] H. Y. Hwang[5,6]

[1]*Bell Laboratories, Lucent Technologies, Murray Hill, NJ 07904, USA*

[2]*Department of Physics and Astronomy, Rutgers University, Piscataway, NJ 08854, USA*

[3]*Mat-Sim Research LLC, Murray Hill, NJ 07904, USA*

[4]*Cornell Center for Materials Research, Cornell University, Ithaca, NY 14853, USA*

[5]*Dept. of Advanced Materials Science, University of Tokyo, Kashiwa, Chiba 277-8561, Japan*

[6]*Japan Science and Technology Agency, Kawaguchi 332-0012, Japan*

Abstract

The atomic-level control achievable in artificially-structured oxide superlattices provides a unique opportunity to explore interface phases of matter including high-density 2D electron gases. Electronic-structure calculations show that the charge distribution introduced by $LaTiO_3$ monolayers in $SrTiO_3$ is strongly modulated by electron-phonon interactions with significant ionic polarization. Anharmonic finite-temperature effects must be included to reproduce experiment. Density functional perturbation theory is used to parameterize a simple model introduced to represent these effects and predict temperature dependencies.

# I. INTRODUCTION

Recent advances in oxide thin film techniques have made possible the growth of artificial superlattices controlled on an atomic-layer scale, opening the way to the exploration of new physical phenomena and potential new electronic and optic device applications.[1,2,3,4] Interfaces between insulators can be intrinsically charged, and constitute new realizations of the two-dimensional electron gas (2DEG) which may display significant differences from conventional semiconductor 2DEG's.[5] Also intriguing are the issues raised by interfaces between qualitatively different materials, such as between a Mott insulator and an ordinary band insulator. Most of the focus on the emerging issue of interface electronic structure in such systems has been placed on electron correlation effects.[6] Here we demonstrate that lattice screening is an effect of central importance. For highly-polarizable ionic materials, including these effects drastically changes the charge distribution. Finite-temperature effects on polarizability through lattice anharmonicity also play a major role. We will introduce an effective and broadly applicable approximation for dealing with this issue.

Our approach is based on the theoretical analysis of a structure examined in a set of recent experiments, an isolated [001] monolayer of $LaTiO_3$ grown in an $SrTiO_3$ host.[7] Substituting a monolayer of trivalent La for divalent Sr effectively introduces a layer of electron-donor dopants, with the extra La electron contributed to conduction-band states of several neighboring $SrTiO_3$ layers to form a 2DEG. The $SrTiO_3$ conduction bands are Ti 3d in character, and while the valence of Ti in the bulk is $Ti^{4+}$, Ti in the 2DEG can be considered to be in a mixed-valent $Ti^{3+}/Ti^{4+}$ state. A scanning transmission electron microscope with near-atomic-scale resolution was used to obtain electron energy loss spectroscopy (EELS) data from core-level excitations as a function of position relative to the La layer. The Ti $L_{2,3}$ near-edge spectra distinctly identify the Ti valence, and spectra for any intermediate valence can be accurately fit by a linear combination of the end-point $Ti^{3+}$ and $Ti^{4+}$ spectra. These fits allowed the charge density to be accurately mapped as a function of distance from the La layer,[7] and we have performed ab-initio electronic structure calculations to understand the mechanisms determining this distribution.

## II. THEORY AND EXPERIMENTAL COMPARISONS

We employed density functional theory (DFT) to perform the calculations reported here. $LaTiO_3$ is considered to be a Mott-Hubbard insulator, and the potential importance of strong electron-correlation effects could render this choice open to question. However, experiments on a series of bulk $Sr_{1-x}La_xTiO_3$ alloys indicate that they behave as Fermi liquids up to x~0.95, with properties consistent with simple rigid-band filling of the $SrTiO_3$ conduction bands up to x~0.8.[8] The maximum effective x measured for the La monolayer system is ~0.3,[7] so we are far from the limit where strong-correlation effects become important for this particular case. Strong-correlation effects with thicker $LaTiO_3$ layers have been examined in a series of studies based on tight-binding models including Hubbard-U onsite interactions.[6] An ab-initio study using the linear muffin tin orbital method in both the LDA and "LDA+U" approximations concluded that the inclusion of the added U interaction didn't make any difference for the single-La-layer case.[9] These studies did not treat the effects of structural relaxation and lattice polarization which are the main focus of the present work.

The majority of the calculations were carried out using a plane-wave basis in adaptive curvilinear coordinates (ACC),[10] norm-conserving pseudopotentials,[11] and the PBE generalized-gradient exchange-correlation functional.[12] The O and Ti pseudopotentials were previously shown to give good structural results for $TiO_2$ with ACC and the same plane wave cutoffs used here (40Ry average, ~130Ry maximum effective).[13] The Sr 4p and La 5p core levels lies near the O 2s in energy and were treated as valence states. We found that to obtain a good lattice constant for bulk $LaTiO_3$, we needed an La f pseudopotential which accurately reproduced the scattering resonance which is the precursor of the partially-filled 4f shells of the lanthanide series. Our calculated values of 3.87Å and 3.90Å are close to room-temperature experimental values of 3.905Å and 3.92Å for the Sr and La compounds, respectively.[14] Most of the calculations utilized a 9-unit tetragonal supercell with space group P4/mmm, the calculated $SrTiO_3$ lattice constant for *a* and *b*, and $c = 34.93Å = (9+\delta)a$, with $\delta = 0.019$ calculated to preserve the (initial) volume of the single $LaTiO_3$ cell. Since the calculated bulk lattice constants differ by only 0.8%, we considered that this procedure was sufficiently accurate that optimization of *c* in the supercell calculation was not warranted. Self-consistency and structural-relaxation calculations utilized 6 **k** points in the reduced Brillouin zone (RBZ).

After self-consistency was achieved for the supercell in the initial structure, its band states were calculated non-self-consistently for a fine mesh of 56 **k** points in the RBZ, and the density of states was calculated. The structure was then relaxed until all forces were <0.0015 eV/Å, and the fine-grid states recalculated. As shown in Fig. 1 for the relaxed structure, the Fermi level $E_F$ lies in the $SrTiO_3$ conduction band. For the unrelaxed case, occupied states are pulled a few tenths of an eV down into the gap. The distribution of the 2DEG among the 9 Ti's was found by summing the densities of each fine-mesh state in a window between mid-gap and $E_F$ within "soft spheres" with mean radii 0.65Å centered at each Ti, and normalizing to one electron. These results for the symmetry-inequivalent Ti's are shown in the first two columns of Table I.

It is immediately clear that the Ti excess charge falls very rapidly with distance from the La layer for the unrelaxed structure, but that it peaks in the center of the supercell in the relaxed case, that is, midway between the La layers. Quantitative comparison of the Table I numbers with experiment requires additional analysis, however. In Figs. 2a and 2b, we have replotted the data points from Fig. 3 of Ref. 7. The data in Fig. 2a are the integrated intensity of the EELS signal of the La $M_5$ edge. Since the annular dark field image in the Fig. 3, Ref. 7 inset image confirms that the La form a single monolayer, the width of the data in Fig. 2a can be taken as the resolution of the EELS measurement. A least-mean-squares fit to this data with a Gaussian is shown in Fig. 2a, with fwhm $9.14 \pm 1.1$Å.

The data in Fig. 2b represent the excess Ti charge. While the cusped-exponential fit to this data shown in Fig. 3b of Ref. 7 appears to be very good, the cusp is not physically reasonable and must be an artifact. First, there are two symmetry-equivalent Ti's at $\pm 1.96$Å relative to the La. Second, this basic 4Å width of the excess charge maximum must be further broadened in the data by the EELS resolution function. Choosing the simplest possible function which will have both the exponential decay seen in the charge tails and the requisite rounding at the origin, we fit the data in Fig. 2b with the function shown. That the "rounding length," 5.76Å, is somewhat larger than the length scale of the resolution Gaussian indicates that these fits are consistent. The exponential decay length in the Fig. 2b fit, 6.91Å, is slightly shorter than that of the cusp fit of Fig. 3b, Ref. 7.

As can be seen from Fig. 2b, there is still enough charge at ±17.5Å that the fit from the isolated experimental La monolayer cannot be directly compared to the 9-layer supercell results. We argue that coherency effects between La layers separated by 35Å should be sufficiently small that overlapping a periodic array of the charge-density fitting functions with the supercell spacing, labeled "synthesized experiment" in Fig. 3, should give an excellent approximation to the charge that would be found were the experimental system actually a 9-layer supercell.

Finally, a periodic array of the discrete Ti charges in Table I must be convolved with the Fig. 2a resolution Gaussian to complete the comparisons between theory and experiment. The unrelaxed decay in Fig. 3 is much faster than experiment. As was already clear from the Table I numbers, the relaxed case is qualitatively inconsistent with experiment. To investigate whether the central peak might possibly be a supercell artifact, we performed a fully-relaxed set of calculations for an 11-layer supercell. These results, in the third column of Table I, show an even more pronounced trapping of the 2DEG in the supercell center. We note in passing that the similarity of polarizations observed for our 9- and 11-layer supercells supports our decision to dispense with optimization of the supercell $c$ lattice constant. Atomic-position relaxation alone with more $SrTiO_3$ layers will increasingly accommodate the small (<<0.8%) possible expansions or contractions of the unit cells adjacent to the $LaTiO_3$ layer.

Examining the ion displacements involved in the relaxation, it is clear that these result in a substantial electric polarization. We indicate this qualitatively in Fig. 4, where the relative displacements of the cations and the anions in each successive [001] plane of alternating SrO and TiO$_2$ composition are plotted. The cations move outwards away from the La layer, which effectively has a charge of +1, while the anions move inwards. These polarizations are apparently too large, overscreen the La, and reduce its ability to bind the 2DEG in an attractive well. The cause for the marked disagreement we have found appears to be the temperature dependence of the SrTiO$_3$ dielectric constant $\varepsilon$ caused by anharmonic vibrational effects. While SrTiO$_3$ does not display a ferroelectric transition, $\varepsilon$ diverges strongly at low temperatures, reaching values $\sim 2\times10^4$.[15] It was characterized as a "quantum paraelectric,"[15] and was later shown to actually undergo a ferroelectric transition at 24K when the quantum mechanical fluctuations were suppressed just slightly by the isotopic substitution of $^{18}$O for $^{16}$O.[16] Since our calculations are effectively zero-temperature while the experiments were performed at room temperature, a substantial difference in the ionic screening was to be expected, although the quantitative effect on the charge distribution was not obvious. The divergence is strictly in the long-wavelength, $\mathbf{q} \to 0$ limit of $\varepsilon(\mathbf{q})$, while our system involves a distribution of finite $\mathbf{q}$ values self-consistently determined by the $\mathbf{q}$ dependence of $\varepsilon(\mathbf{q})$ and the charge decay length.

To verify that excessive polarization as opposed to some other artifact of our calculation was the root cause of the supercell center peak, we calculated the Ti electron distribution for an ad hoc geometry formed by averaging our relaxed and unrelaxed structures for a 9-layer system. These "half-relaxed" results are given in the fourth column of Table I. This charge monotonically decays toward the supercell center, but a quantitative comparison with the data, not shown, indicates that the decay is considerably too rapid. Further refinement is unwarranted as such a fit is not simply related to any measurable experimental property.

### III. APPROXIMATE FINITE-TEMPERATURE THEORY

An approach to the finite-temperature effects is suggested by the self-consistent phonon approximation.[17] While ab-initio calculation of all the anharmonic potentials needed to calculate this temperature-dependent variational harmonic approximation is not practical, the concept is useful and can be adapted to our purposes. Density functional perturbation theory (DFPT) permits the direct calculation of the electronic and ionic contributions to $\varepsilon$.[18] In particular, the calculation of the ionic contribution proceeds by calculating the interatomic force constants at $\mathbf{q} = 0$ and the Born effective charge tensors, and combining them appropriately.[19] Our approach consists of 3 steps. (1) Calculate the ab-initio force constants and charge tensors for $SrTiO_3$, which of course gives T=0 results. (2) Introduce a set of supplementary harmonic forces intended to simulate the anharmonic stiffening. (3) Adjust a parameter of these forces to fit the measured room-temperature $\varepsilon$.

We used the ABINIT open-source electronic structure code[20] to perform our DFPT calculations. This code uses a simple plane-wave basis set, so in using it with our pseudopotentials we set the cutoff in the range of our maximum effective ACC cutoff. We found that there were in fact unstable polar optical phonon modes at $\mathbf{q}$=0, which is physically correct considering that the low-temperature system is stabilized by quantum fluctuations. For our supplementary force model, we introduced a set of identical springs between all near-neighbor anion-cation pairs. While a range of values are reported in the literature, we chose to fit the single-crystal value $\varepsilon = 290$ at T=298K.[21] Second energy derivatives corresponding to our model were patched into the ABINIT calculation, and a trial-and-error search yielded the spring-constant value 49.6 eV/Å$^2$. Recognizing that we are "tuning" this system near a singularity (in the inverse force-constant matrix), we tested convergence with ab-initio force constants and charges calculated at 100, 120 and 140Ry. The results were $\varepsilon = 245$, 293 (fit), and 286, which we regard as quite satisfactory. We note that a study using the local density approximation found the cubic phase to be stable at T=0, and in fact to have a T=0 $\varepsilon$ close to the T=298K experimental value. [22] These authors recognized this as an accidental consequence of the smaller LDA lattice constant. Because of the "critical tuning" nature of the dielectric anomaly, LDA results for the La-layer-induced polarization could be quite different from GGA. Based on the Ref. 22 results, a zero-temperature LDA calculation for the Ti charge distribution would likely give results close to experiment. However, this does not represent the correct physics, and would conceal the role played by temperature and anharmonicity.

Introducing the springs into our 9-layer supercell (including the same spring between La and its nearest O neighbors), we recomputed the structural optimization and found the excess charge distribution shown in the last column of Table I. These results, convolved with the resolution function, are plotted in Fig. 3 and the agreement with experiment excellent. The polar displacements are shown in Fig. 4, and are closer to the fully-relaxed results than we might have anticipated. This illustrates the $\varepsilon(\mathbf{q})$ effects, namely that the difference between room-temperature and low-temperature polarization relevant to the electron charge decay does not involve as large an enhancement factor as does the macroscopic $\varepsilon(0)$, i.e. $\sim 10^2$. Anharmonic effects are predominantly local, and our "supplementary spring" model, despite having been fit to a **q**=0 datum, entails local forces and can represent finite-**q** effects.

Since they represent anharmonic effects, the spring constants of our supplementary springs are temperature dependent, to be fitted to the measured macroscopic $\varepsilon$ at each temperature at which we wish to compute the La-induced polarization. The results of Ref. 22 imply that an LDA calculation would need negative force constants to compare to data taken below room temperature. While an LDA+U functional did not change the charge distribution results from LDA for the unrelaxed single-La-layer case,[9] we anticipate that it would require yet another set of tunings of the spring constants at various temperatures.

## IV. SUMMARY AND DISCUSSION

In conclusion, we have identified the lattice polarization and anharmonic contributions to its temperature dependence as very important physical effects governing the charge profile in oxide superlattices. Calculations for another type of oxide 2DEG, one confined in a quantum well formed from $SrTiO_3$ and a 4-cell thick layer of the wider-gap band insulator $LaAlO_3$, show similar large polarization effects.[23] Our approximate method for treating finite-temperatures should be broadly applicable, and the temperature dependence of charge confinement should be calculable wherever macroscopic dielectric data is available over the temperature range of interest. The variability of the effective confinement potential will necessarily influence transport and magnetotransport properties as well. While we did not need to simultaneously treat strong electron-electron correlation effects and strong electron-polarization coupling for the case of the single LaTiO$_3$ layer, charged interfaces involving thick Mott insulators will require both.

Note added: After the completion of this work, we became aware of a recent independent zero-temperature treatment of lattice polarization for LaTiO$_3$-SrTiO$_3$ heterostructures.[24]

## ACKNOWLEDGEMENTS

The authors thank David Vanderbilt for helpful discussions. D. M. is supported by IRG-3 of Cornell's Center for Materials Research, an NSF MRSEC.

**Tables**

| Ti | UNRLX | RLX | RLX-11 | RLX/2 | "298K" |
|---|---|---|---|---|---|
| 1 | 0.4248 | 0.1691 | 0.1258 | 0.3129 | 0.2348 |
| 2 | 0.0621 | 0.0902 | 0.0699 | 0.0962 | 0.1070 |
| 3 | 0.0108 | 0.0870 | 0.0670 | 0.0446 | 0.0683 |
| 4 | 0.0019 | 0.0974 | 0.0874 | 0.0311 | 0.0596 |
| 5 | 0.0007 | 0.1124 | 0.0949 | 0.0303 | 0.0607 |
| 6 |  |  | 0.1100 |  |  |

**Table I.** Excess electron charge per Ti, with Ti's numbered sequentially with increasing distance from the La layer. Column headings abbreviate unrelaxed, relaxed, 11-layer relaxed, "half-relaxed," and simulated T=298K results, respectively.

# Figure Captions

Fig. 1. Density of states for the 9-layer fully-relaxed supercell, with inset showing valence and conduction band edges.

Fig. 2. (color online). Original data from Ref. 7 with fitting functions shown for (a) the La EELS intensity profile and (b) the Ti excess charge. Fit functions are for $x$ in Å units.

Fig. 3. (color online). Comparison of calculated and experimental results. "Synthesized experiment" indicates periodically-overlapped fit functions.

Fig. 4. (color online). Structure and displacement of cations and anions from coplanarity for alternating (La/Sr)O and $TiO_2$ layers. Positive values indicate that cations move to the right relative to the O anions.

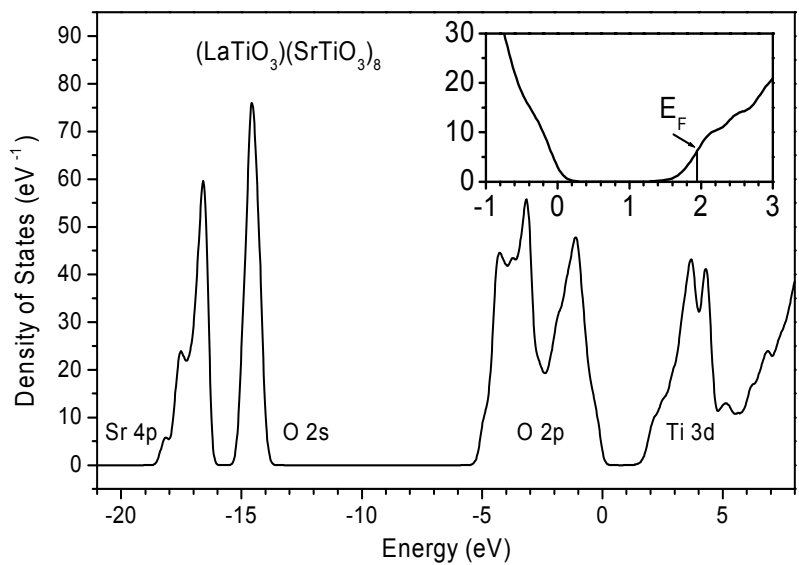

Fig. 1.

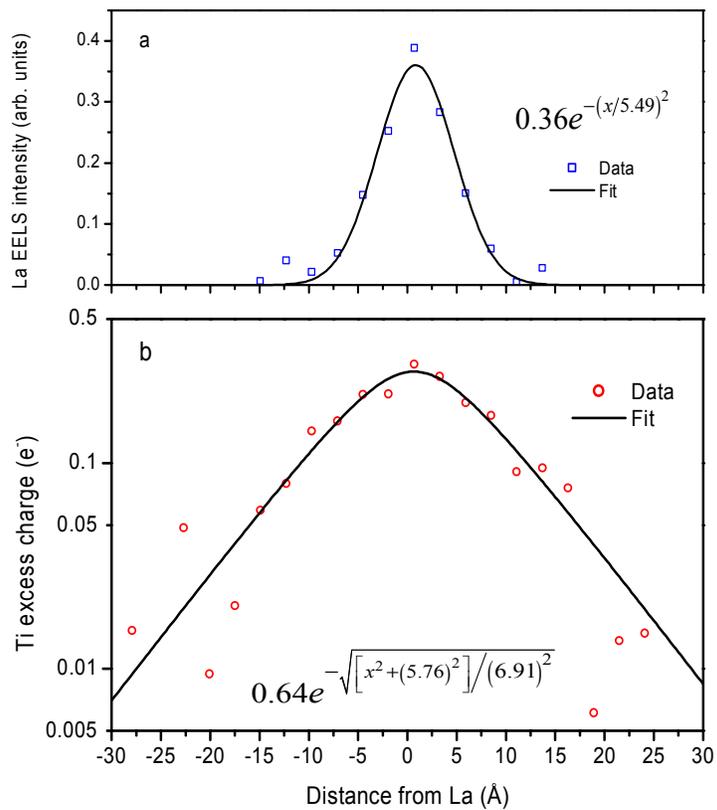

Fig. 2

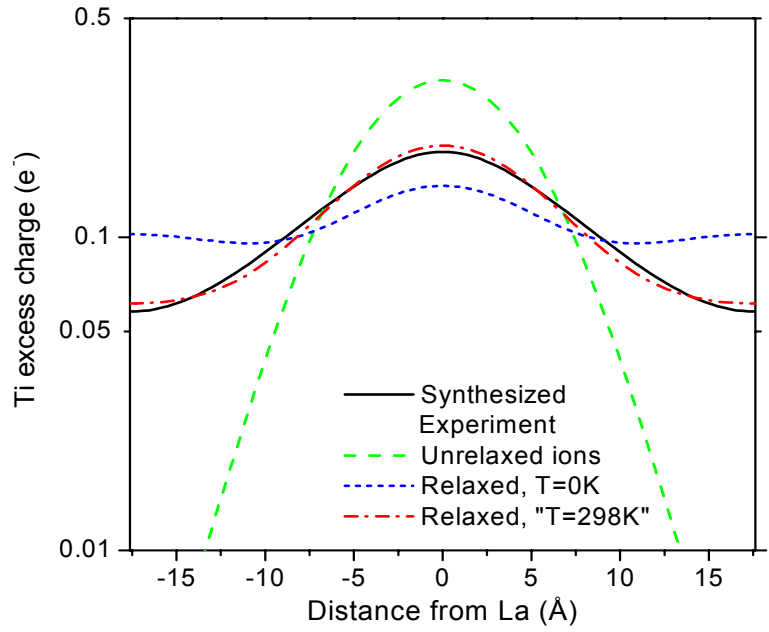

Fig. 3

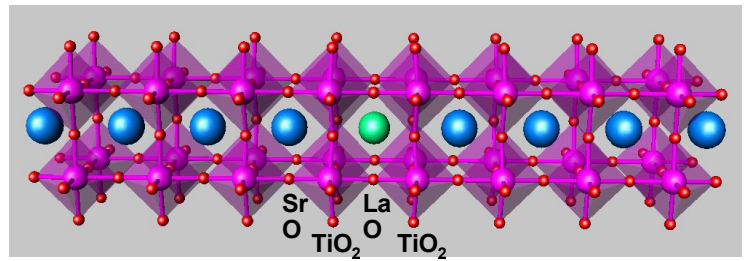

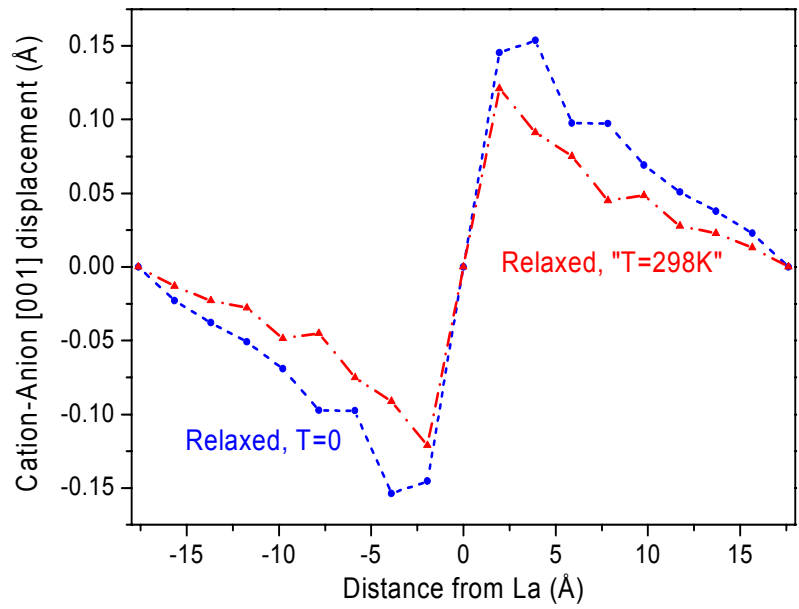

Fig. 4